\documentclass[conference]{IEEEtran}

\usepackage[utf8]{inputenc}
\usepackage[T1]{fontenc}

\ifCLASSINFOpdf
  \usepackage[pdftex]{graphicx}
  \graphicspath{{figures/}}
  \DeclareGraphicsExtensions{.pdf,.jpeg,.png}
\else
  \usepackage[dvips]{graphicx}
  \graphicspath{{figures/}}
  \DeclareGraphicsExtensions{.eps}
\fi
\begin{document}
\title{Automated Application Offloading through Ant-inspired Decision-Making}

\author{\IEEEauthorblockN{Roya Golchay\IEEEauthorrefmark{1},
Frédéric Le Mouël\IEEEauthorrefmark{2}, Julien Ponge\IEEEauthorrefmark{3} and
Nicolas Stouls\IEEEauthorrefmark{4}}
\IEEEauthorblockA{Univ Lyon, INSA Lyon, CITI, F-69621 Villeurbanne, France\\
Email: \IEEEauthorrefmark{1}roya.golchay@insa-lyon.fr,
\IEEEauthorrefmark{2}frederic.le-mouel@insa-lyon.fr,
\IEEEauthorrefmark{3}julien.ponge@insa-lyon.fr,
\IEEEauthorrefmark{4}nicolas.stouls@insa-lyon.fr}}

\maketitle

\begin{abstract}
The explosive trend of smartphone usage as the most effective and convenient communication tools of human life in recent years make developers build ever more complex smartphone applications. Gaming, navigation, video editing, augmented reality, and speech recognition applications require considerable computational power and energy. Although smartphones have a wide range of capabilities - GPS, WiFi, cameras - their inherent limitations - frequent disconnections, mobility - and significant constraints - size, lower weights, longer battery life - make difficult to exploiting their full potential to run complex applications. Several research works have proposed solutions in application offloading domain, but few ones concerning the highly changing properties of the environment. To address these issues, we realize an automated application offloading middleware, ACOMMA, with dynamic and re-adaptable decision-making engine. The decision engine of ACOMMA is based on an ant-inspired algorithm.
\end{abstract}


% \fi
%\IEEEpeerreviewmaketitle

\section{Introduction}

The explosive trend of smartphone usage as the most effective and convenient communication tools of human life in recent years - with the 50 percent growth rate in 2013~\cite{Cis} - make developers to build ever more complex smartphone applications such as gaming, navigation, video editing, augmented reality, and speech recognition, which require considerable computational power and energy.

However smartphones have a wide range of capabilities, typically including GPS, WiFi, cameras, gigabytes of storage, and gigahertz-speed processors, the importance and desirability of smaller sizes, lower weights and  longer battery life as well as their inherent limitations such as resource scarcity, frequent disconnections and mobility, make them difficult to exploiting their full potential to run these complex applications and have the best performance.

It seems to keep pace with increasing performance requirements, mobile users have to continually upgrade their hardware to augment the computational power as applications become more complex but still experience some limitations specially short battery lifetime. More feasible approach is empowering mobile devices using software solutions-application offloading, that improves the performance and the energy consumption of resource-poor mobile devices by using the power of one or more resource-rich stations. A key area of application offloading is to apply a remote execution of an application - totally or partially - to resource-intensive devices to improve performance and energy consumption. The surrogate can be a powerful stationary device or a set of processors. Drastic evolution of wireless technologies that make network connectivity ubiquitous and successful practices of Cloud Computing for stationary machines are motivating factors to bring the cloud to the vicinity of a mobile from an offloading perspective. As a result, Mobile Cloud Computing was introduced to enable rich mobile computing by extending the on-demand computing vision of Cloud Computing and enrich smartphones and address their issues of computational power and battery lifetime by executing complete mobile applications or identified resource intensive components of a partitioned mobile application on the cloud-based surrogates~\cite{Fernando2013}. Several research works have proposed solutions~\cite{Cuervo:2010}, \cite{Kosta2012} but proposed partitioning mechanisms are not appropriate to highly changing environments.

%Considerable research work have proposed solutions in application offloading domain. MAUI~\cite{Cuervo:2010} and ThinkAir~\cite{Kosta2012} are among the well-known approaches for application offloading which their goal is to optimize the energy consumption or the application execution time using VM migration. We believe that this virtualized environment is heavy for small mobile devices. CloneCloud~\cite{Chun:2011} is another approach which splits application to make offloading but its partitioning mechanism is off-line using a static analyzer. Concerning the highly changing environment of Mobile Cloud Computing, what might be the right partitioning at a time, may be the wrong one later. In addition, for all these approaches special development of application or code annotation is required.

To address these issues, we propose ACOMMA, an \textbf{A}nt-inspired \textbf{C}ollaborative \textbf{A}pplication \textbf{O}ffloading \textbf{M}iddleware for \textbf{M}obile \textbf{A}pplications. ACOMMA is an automated application offloading middleware with dynamic and re-adaptable decision-making engine based on an ant-inspired algorithm. To take more flexible offloading decisions, ACOMMA performs fine-graineded method-level application partitioning. Our proposed middleware, by dint of its service-based architecture, could be used by devices that support web services without any special requirements. This middleware is also equipped with a learning-based decision-making process to avoid running a complete decision-making process in duplicate situations. To eliminate the role of developer for special development of mobile application or its annotating for application partitioning, we propose an application transformer to modify the application into an adaptable form with this offloading middleware.

The underlying motivation for ACOMMA lies in the following intuition: however Mobile Cloud Computing is very beneficial, there are still some challenges arise from mobile devices, clouds and their interactions. Meanwhile addressing communication and application development complexity challenges, ACOMMA focuses on offloading performance. It applies an ant-inspired algorithm to perform fine-grained and method-level application offloading considering two performance criteria at the same time. In short, our main contributions consist in:
\begin{itemize}
\item designing and developing an open and service-oriented architecture that makes ACOMMA adaptable to devices that support web services without special API requirements,
\item making ACOMMA flexible with any Android mobile application due to its application transformer,
\item designing an automated offloading middleware that dynamically makes efficient offloading decisions considering the changing environment by using bi-objective algorithm,
\item adding a learning feature in the decision-making process to avoid re-execution of decision-making algorithms when offloading decisions have already been taken in similar situations.
\end{itemize}

In what follows, we first explain in Section~\ref{section:transformation} how a mobile application is automatically modified by our application transformer to be able to be offloaded by ACOMMA. Then we present the design of the architecture of ACOMMA in Section~\ref{section:architecture} and its decision-making engine that decides of the offloading based on an ant-inspired algorithm in Section~\ref{section:decisionmaking}. We describe our implementation and experimental evaluations of the prototype in Section~\ref{section:evaluations}. We survey related work in Section~\ref{section:stateoftheart}, and finally conclude and discuss limitations in Section~\ref{section:conclusion}.

\section{Application Transformation}
\label{section:transformation}

Regardless of the current situation in the mobile applications, there may exist some components that must be executed locally because of their inherent dependencies to the mobile device. In almost all existing offloading middlewares, even in the ones that take online offloading decisions, the developer is responsible in detecting and annotating these components. Its is clear that the quality of offloading is highly dependent on knowledge, expertises, and experience of the developers who annotate applications. Any small issue in annotating may cause big changes in the offloading process. To eliminate any need for manual annotation and modification, we propose an application transformer that detects offloadable parts of the application by applying some rules.

To be adaptable with a client-server architecture and supporting REST/HTTP communication in its service-oriented model, this application transformer also has the responsibility of code modification. The application methods which are offloadable parts in our method-level application offloading, should be modified in a form where methods act as services and are accessible via REST. Servicizing is what we call this changing process.

\begin{figure}[!htb]
\centering
\includegraphics[width=\columnwidth]{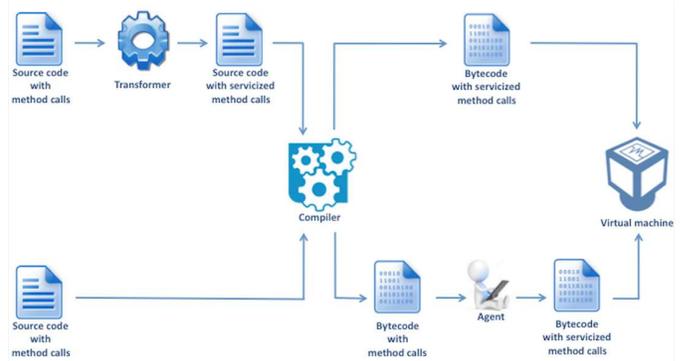}
\caption{Mobile application execution flow with servicization modifications}
\label{fig:transformer}
\end{figure}

The transformer picks the application as the input and creates the modified application in output. Furthermore, to be located on the server, a version of the application is provided that adds the accountability to services for any method. Depending on the circumstances, this transformer can be run directly on the mobile device or on any other machine, and the modified application in output is transferred to the mobile device for executing.

Figure~\ref{fig:transformer} illustrates the flow of a mobile application execution with the transformer. The application transformer gets the source code and generates a new version of it with offloadable servicized method calls. This new source code is transformed to bytecode by the compiler and the virtual machine interprets the stream of bytecode as a sequence of instructions and then executes it to produce desired output.

To supports applications without open source codes, we add an agent to the execution flow that transforms the bytecode into a sevicized bytecode before the interpreting by the virtual machine. Although the ability to change the bytecode makes the approach more general and dynamic, it complicates the modification process and reduces the efficiency because the agent must be present on the mobile device. The availability of the source code, however, allows the application transformation process to be done on a system other than the mobile device and the bytecode of the modified application to be installed on the mobile device. Having the development chain before the mobile device increases the efficiency and performance, but decreases the dynamism while the system is no longer able to exert next changes during the execution.

\section{The architecture of ACOMMA}
\label{section:architecture}

\begin{figure}[!htb]
\centering
\includegraphics[width=\columnwidth]{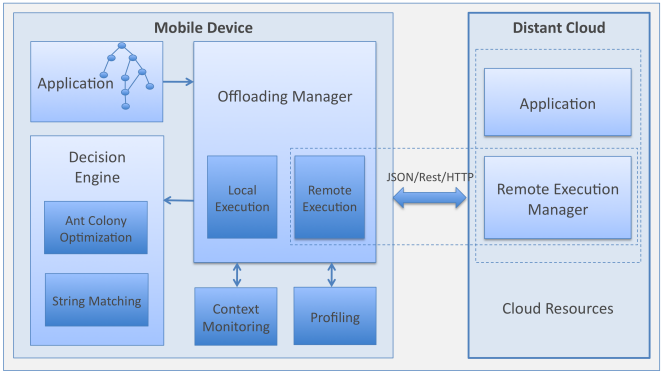}
\caption{An architectural view of offloading building blocks in ACOMMA}
\label{fig:architecture}
\end{figure}

ACOMMA has an open and service-oriented architecture with an offloading service, a context monitoring service and a profiling service as building blocks. This open architecture with REST as the communication API makes ACOMMA generic and usable for the devices that support web services. The offloading service is responsible for partitioning a mobile application in a dynamic way and performs offloading. The context monitoring and profiling services collect the required data for the offloading service. They provide environmental information such as application type, cloud information and communication conditions as well as user information such as his requirements, preferences, and limitations. As shown in Figure~\ref{fig:architecture}, the offloading service gets a mobile application as an input, employs the collected data of context monitoring and profiling services, determines offloading by the help of its decision-making algorithm in decision engine, especially which parts of the application lead to higher performance, and finally performs offloading.

We consider a mobile application modelled as a graph where vertices and edges represent methods and their dependencies in term of method calls respectively. In such a call graph, the graph partitions represent the executing environments of partition members. %For instance in sample graph of figure ~\ref{fig:callGraph}  there is just one cut that breaks apart two partitions where all methods execute locally except methods b and e that execute remotely on a distant execution environment.

The decision engine is in charge of partitioning the call graph in an efficient way to determine offloadable parts of the application. Unlike existing offloading middlewares that use linear programming to cut a graph based on a single objective, we are interested in taking into account two criteria at the same time for graph partitioning. This bi-criteria decision-making process helps ACOMMA to perform more dynamic and flexible offloading concerning the highly changing environment. To take such a decision, as the constraints in dynamic environments can never be guaranteed to an optimal solution, we need to apply heuristic approaches like genetic algorithms, fuzzy logics or bio-inspired algorithms. Because of their collaborative decision-making process, self-organization, autonomy, and vigor for solving optimization problems, we use bio-inspired algorithms and especially an ant-inspired algorithm that is robust, self-organized and flexible.

\section{Dynamic Decision-Making}
\label{section:decisionmaking}

\subsection{Shortest Path Problem}

\begin{figure}[!htb]
\centering
\includegraphics[width=\columnwidth]{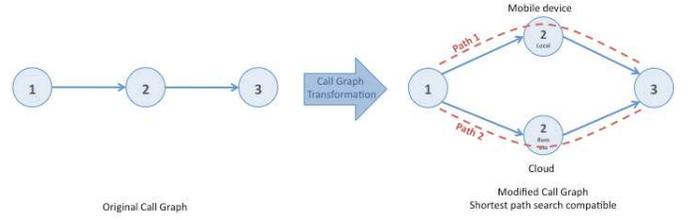}
\caption{Modifying  call graph to be compatible to Shortest Path problem}
\label{fig:callGraph-transform}
\end{figure}

Ant-inspired algorithms are used to solve different problem types, but they are more adapted to Shortest Path Problems. We propose the partitioning problem investigated as a Shortest Path Problem where the application call graph is modified in a way that its nodes belong to local and remote executing environments. In the graph modification process, as shown in Figure~\ref{fig:callGraph-transform}, all graph nodes are duplicated instead of having only one single node and consequently their coupling vertices. As the nodes represent the methods, the original nodes show methods on the mobile device and the duplicated ones refer to the corresponding methods on the cloud.

In the transformed graph, choosing \emph{Path1} shows the local execution of method2, where \emph{Path2} represents its remote execution in the cloud. The goal of Shortest Path Problems is finding a path between two nodes in a weighted graph such that the sum of the weights of its constituent edges is minimized. As we want to take bi-criteria offloading decisions, there are two attributed weights for each edge.
We consider CPU usage and execution time as constraints of the decision-making and aim to find an offloading solution to minimize both of them. For the same mobile device and cloud, any change in the network conditions directly affects the execution time, for example, more network load leads to an increase in the execution time. There is also a direct relationship between the CPU usage and the battery consumption, the more an application uses CPU power, the more it consumes battery. Although it seems that the applied criteria are just the execution time and CPU usage, the network communication conditions and the amount of battery consumption also influence the offloading decision-making process.

\begin{figure}[!htb]
\centering
\includegraphics[width=\columnwidth]{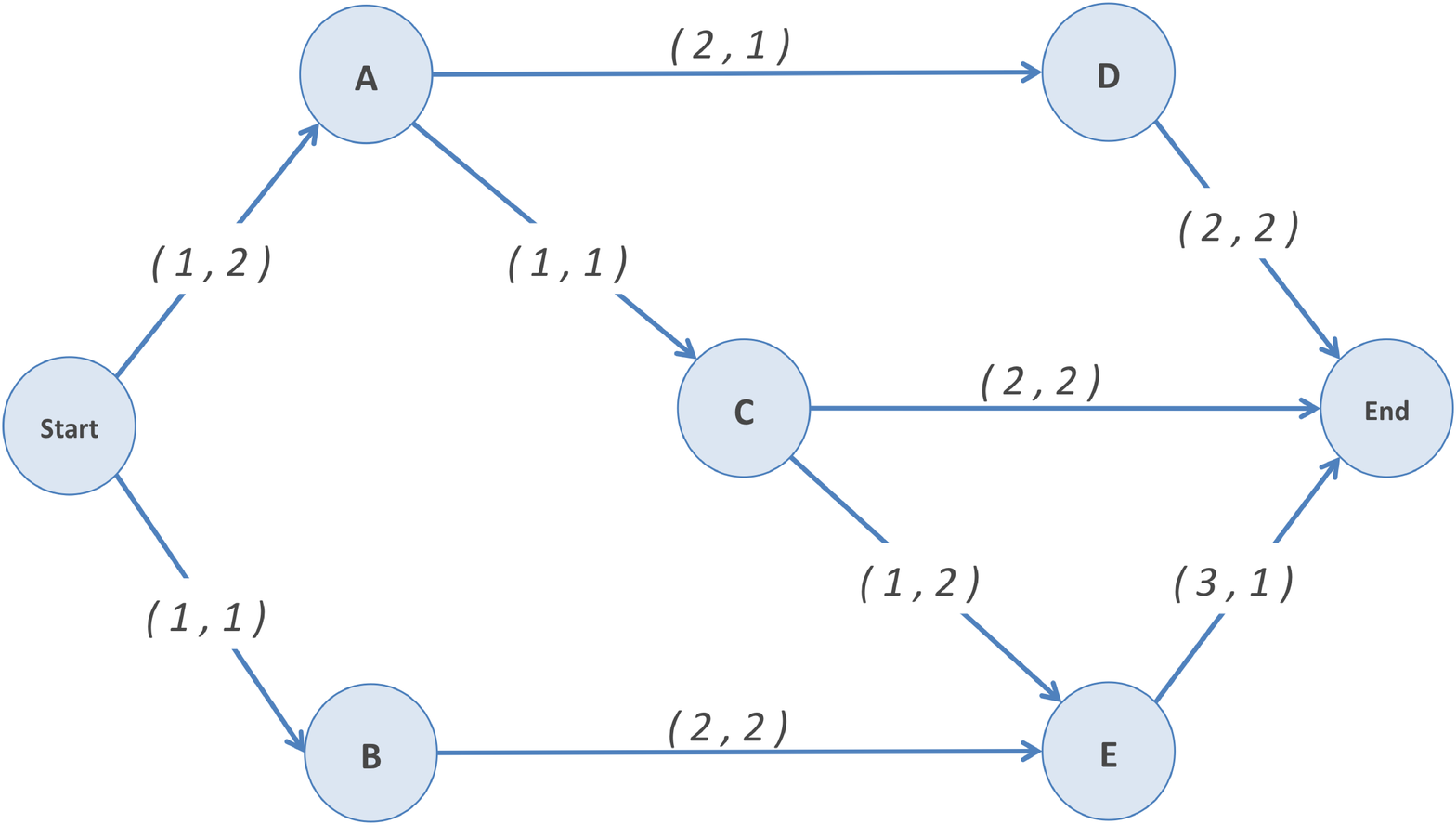}
\caption{Solution set for Shortest Path Problem :(4,5),(6,4)}
\label{fig:non-dominated}
\end{figure}

Unlike single-objective optimization problems resulting in a scaler optimal solution, solving bi-objective optimization problems concentrate in finding a tradeoff between two objectives and resulting in a set of solutions called non-dominated solutions. For example, in Figure~\ref{fig:non-dominated}, there exist four paths between the start and the end points: 'start-A-D-end', 'start-A-C-end', 'start-A-C-E-end' and 'start-B-E-end' with their respective related objective functions of (5, 5), (4, 5), (6, 6) and (6, 4). In this graph, the objective functions (5,5) and (6,6) are dominated by (4,5), however between (4,5) and (6,4) the best path cannot be chosen because none is dominated by the other. As a result, the non-dominated paths are 'start-A-C-end' and 'start-B-E-end' with (4,5),(6,4) as objective functions.

Based on our knowledge, this is the first time that an application partitioning problem in offloading decision-making process is considered as a Shortest Path Problem. In the following, we explain how ACOMMA utilizes Ant Colony Optimization algorithm to solve the Shortest Path Problem.

\subsection{Decision-making using Ant Colony Optimization}

In nature, an ant starts looking for food in a random manner. After having found the food, they make more or less directly to the nest, depositing an odorous called pheromone. Thanks to this trace, other ants can come to find food. Over time, the pheromone trail starts to evaporate. The more time it takes for an ant to travel down the path and back again, the more time the pheromones have to evaporate. A short path, by comparison, gets marched over more frequently, and thus the pheromone density becomes higher on shorter paths than longer ones.

We apply~\cite{Ghoseiri2010} as Ant Colony Optimization algorithm~(ACO) in which two pheromone matrices are intended for two objectives. These pheromone matrices are updated at the end of each iteration separately based on the generated results. In addition, when an ant moves from one node to another, the pheromone trail is locally updated according to the evaporation rate. An artificial ant moves from one node to the next one based on a series of transition rules and with the help of two heuristic parameters.

When an application starts, as an initialization phase, ACOMMA creates application's service graph in an aforementioned form where both costs of CPU usage and execution time of each method are set to zero. For the first method call, when the algorithm executes for the first time, the non-dominated set consists of all possible paths. ACOMMA randomly selects one of these solutions and execute the application. Then, weights of all edges of selected path are updated with real values of each method execution costs. So, during the next algorithm execution, this selected path as well as the paths within common edges won't be in the non-dominated set. After several executions, all graph vertices will have real weights progressively, and Ant Colony Optimization algorithm by applying local and global updates of pheromone trail, gives different non-dominated paths.

To prevent the algorithm execution for duplicate situations, we establish a learning-based decision-making process that uses the previous decisions made using the Ant Colony Optimization algorithm for the same application and the same situation. ACOMMA saves the history of each application run as a string of executed methods and their execution platform in a cache and applies a simple string matching algorithm, to find the appropriate execution string in this history. The next section shows the evaluations of the decision-making process of ACOMMA.

\section{Implementation and Evaluations}
\label{section:evaluations}

\subsection{Benchmark applications and Experimental platforms}

To evaluate the performance of ACOMMA, we start with four micro-benchmarks and extend our tests with two macro-benchmarks that are representative of popular applications. 

As micro-benchmarks, we develop some mathematical functions of Fibonacci, Matrix multiplication, Matrix determinant, and Integrate. Although these functions are short and simple, they are different enough to allow us to do a variety of tests at the first step. Fibonacci and Matrix multiplications are both composed of a few number of methods but with different mathematical complexities. Fibonacci repeats a basic mathematical operation many times where matrix multiplications do some more complicated calculations. Determinant and Integrate have some number of methods that could be offloaded where the determinant works recursively. Varying the inputs of each of these functions leads to interesting results.

Our macro-benchmarks are Monte Carlo and Face recognition algorithms. Monte Carlo algorithm is a randomized algorithm whose running time is deterministic, but whose output may be incorrect with a certain (typically small) probability. This algorithm could be used for the choice of the next move in a chess game. The Face recognition algorithm tries to match a given face image to a set of given face images using a number of eigenfaces~\cite{face} and is representative of image processing applications.

We use a MacBook Pro with 8 GB of memory and a 2,53 GHz Intel processor dual-core as our remote server. This server has OS X 10.9.5 Mavericks as operating system. We use two different mobile clients to evaluate the decision-making process. The first one is a Samsung Galaxy SII with 1,2 GHz dual-core processor and 2GB of memory running Android version 4.1.2 (Jelly Bean). The second is an Asus Google Nexus 7 Tablette with quad-core 1.2 GHz processor and 1GB of memory running Android version 5.1.1 (Lollipop).

To successfully validate ACOMMA, we need to show that ACOMMA is able to make a correct and efficient offloading decision to improve application performance by selecting an appropriate execution path on the application service graph using the Ant Colony Optimization algorithm. It may also ameliorate its performance while making offloading decisions benefiting from its string matching algorithm.

\subsection{Results}

%\begin{table*}[!htb]
%\centering
%\begin{tabular}{||c|c||c||c||c||c||c||}
%\hline
%             &          &   &     &     &  Face  &\\
%         &  Fibonacci &  Multiplication & Determinant &   Integrate &  Recognition & Monte Carlo \\
%
%\hline
%Serie 1 & 500 & 50x50 & 2 & 1.0-1.5 & 100000x1 & 10-5 \\
%\hline
%Serie 2 & 1000 & 60x60 & 3 & 1.0-2.0 & 100000x2 & 20-7 \\
%\hline
%Serie 3 & 1500 & 70x70 & 4 & 1.0-2.5 & 100000x3 & 30-9 \\
%\hline
%Serie 4 & 2500 & 80x80 & 5 & 1.0-3.0 & 100000x4 & 40-11 \\
%\hline
%\end{tabular}
%\vspace{5pt}
%\caption{Different inputs series for decision-making using ACO}
%\label{inputs}
%\end{table*}

\begin{table*}[!htb]
\centering
\begin{tabular}{|c|c|c|c|c||c|c|c||c|c|c||c|c|c||c|c|c|}
\hline
         & & \multicolumn{3}{c||}{Fibonacci} &  \multicolumn{3}{c||}{Multiplication} &  \multicolumn{3}{c||}{Determinant} &   \multicolumn{3}{c||}{Integrate} \\
\hline
                 & & Success & Time & CPU  &  Success  & Time & CPU &  Success & Time  & CPU &  Success & Time & CPU \\
                 & &  ($\%$) & ($\%$) &  ($\%$) &  ($\%$) & ($\%$) & ($\%$)& ($\%$) & ($\%$) & ($\%$) & ($\%$) & ($\%$) &  ($\%$) \\
\hline
Galaxy S2& Serie 1 &
60 & 40.54    & 20  &
48 & 7.6 & -16.66 &
72  & 31.43 & 66.67  &
96 & 93.97 & 86.55 \\

         & Serie 2 &
64 & 29.7  & 18.75  &
60 & 11.93 & 0 &
100 & 66.20 & 93.73  &
96 & 96.99 & 99.49 \\

         & Serie 3 &
52 & 27.25 & 15.38  &
 52 & 13.06 & 0 &
100 & 92.82 & 97.36  &
96 & 97.78 & 99.55 \\

         & Serie 4 &
56 & 31.08 & 7.14   &
52 & 8.37 & 15.38 &
100 & 98.23 & 99.54  &
96 & 98.59 & 99.81 \\

\hline
Nexus 7 Tablette & Serie 1 &
76 & 13.34 & 10.53  &
48 & 9.74 & -8.33 &
88  & 32.88 & 30.30  &
96 & 96.84 & 98.44 \\

                 & Serie 2 &
52 & 18.98 & 23.08  &
44 & 7.03 & 0 &
96  & 83.78 & 82.97  &
96 & 98.39 & 98.65 \\

                 & Serie 3 &
80 & 24.38 & 53.67  &
56 & 4.30 & -7.14 &
96  & 96.23 & 98.53  &
96 & 98.85 & 99.18 \\

                 & Serie 4 &
56 & 13.23 & 13.69  &
48 & 3.72 & 8.33 &
100 & 98.86 & 99.30  &
96 & 99.15 & 99.55  \\

\hline
\end{tabular}
\vspace{5pt}
\caption{Summary of individual decision-making using ACO on micro benchmarks}
\label{micro-decision-results}
\end{table*}

\begin{table*}
\centering
\begin{tabular}{|c|c|c|c|c||c|c|c||}
\hline
         & & \multicolumn{3}{c||}{Face Recognition} &  \multicolumn{3}{c||}{MonteCarlo} \\
\hline
                 & & Success & Time & CPU  &  Success  & Time & CPU \\
                 & &  ($\%$) & ($\%$) &  ($\%$) &  ($\%$) & ($\%$) & ($\%$) \\
\hline
Galaxy S2 & Serie 1 & 100 & 83.92 & 81.43 & 96 & 81.66 & 99.74 \\
         & Serie 2 & 100 & 87.89 & 60.47 & 100 & 83.72 & 97.74\\
         & Serie 3 & 100 & 88.52 & 75.23 & 96 & 94.68 & 99.92 \\
         & Serie 4 & 100 & 89.98 & 77.89 & 96 & 96.05 & 99.88 \\
\hline
Nexus 7 Tablette & Serie 1 & 96 & 83.74 & 89.68 & 96 & 95.28 & 96.73\\
                 & Serie 2 & 92 & 82.09 & 91.71 & 100 & 94.39 & 97.36\\
                 & Serie 3 & 100 & 80.77 & 84.82 & 96 & 98.64 & 99.15 \\
                 & Serie 4 & 96 & 78.40 & 75.42 & 96 & 99.01 & 99.34\\
\hline
\end{tabular}
\vspace{5pt}
\caption{Summary of individual decision-making using ACO on macro-benchmarks}
\label{macro-decision-results}
\end{table*}

To evaluate the decision-making process of ACOMMA, we run several tests on each benchmarks and we compare the total execution time and CPU usage of an application execution when it executes locally on mobile devices with its execution when offloaded by ACOMMA. To be able to compare the offloading gain in different execution complexities, we run each application 25 times for each and with different inputs (Series 1-4: Fibonnaci 500 to 1500, Multiplication 50x50 to 80x80, Determinant 2 o 5, Integrate 1.0 to 3.0, Recognition 100000x1 to x4, Monte Carlo 10-5 to 40-11). %Inputs of micro- and macro-benchmarks are listed in Table~\ref{inputs}.

Since the execution time and CPU usage of application methods are decision-making criteria that Ant Colony Optimization uses for graph weights while cutting it, we show the gain regarding these parameters while offloading. As expected for solutions of the bi-objective optimization problem, the results show a similar form of gain for both execution time and CPU usage. However, Fibonacci and Matrix multiplications gain in terms of execution time and CPU usage while offloaded by ACOMMA, the gain of Determinant and Integrate is much higher. Fibonacci and matrix multiplication use simple calculations that do not consume considerable resources. In addition, their consumption growth rate is very small. So offloading is less efficient for them compared with more consuming applications and even in some runs, offloading execution takes more time than local execution. Contrariwise, Integrate and Matrix determinant are consuming benchmarks with a significant consumption growth rate as input changing. Using ACOMMA the most consuming parts of the application execute on the server and while the execution time of these parts with different inputs on the server is almost the same, the total execution time using offloading is in the same range while the local execution time grows exponentially with more consuming inputs. In fact, the more the application is resource consuming, the more we gain using offloading.

Summary Tables~\ref{micro-decision-results} and~\ref{macro-decision-results} show the success rate, time gain and CPU gain of ACO while applying on different applications running on different devices for micro-benchmarks and macro-benchmarks respectively.

The successful runs are the runs with their offloading execution time less than their local execution time, in other words, a run is successful if it gains in terms of execution time while offloading. The average success rate of Fibonacci and Matrix multiplication is 62\% and 59.5\% respectively while they augment to 94\% and 96\% for Determinant and Integrate. For Fibonacci and Matrix multiplication, the gain in CPU and time is at 99\%.
Coming to macro-benchmarks, the results show that the gain for Face recognition and Monte Carlo is less than for Integrate and Determinant, but remains really significant with an average success ratio of 98\% and 97\% but with a gain in time and CPU less important of 78 to 89\%. These applications are all consuming but Face recognition and Monte Carlo have a larger service graph that needs more time to find non-dominated solutions. It seems that the efficiency of ACOMMA, depends on the graph complexity as well as the resource consumption of its nodes.

Although the overhead of Ant Colony Optimization algorithm of 10\% in terms of execution time is quite low compared to its gain, we apply a simple string matching algorithm to verify if passing through the paths that are already determined by Ant Colony Optimization algorithm in previous runs is beneficial and can lower this overhead. To this end, the already passed paths are saved in a cache. In the next runs, ACOMMA searches for matches in the cache firstly, and if not found, it runs Ant Colony Optimization algorithm. We have tested string matching without cache invalidation and with it. We applied periodically cache invalidation based on predefined run numbers.
The results show that by using string matching, the total execution time is only slightly improved by 2.3\%, but the overall decision-making overhead is reduced to 5 to 7\%. There is also no big differences between string matching improvements with and without cache invalidation. They may happen for more complex applications with larger service graphs that imply a larger cache in terms of both path size and number of paths.
The results also show that for Face recognition and Monte Carlo algorithms, string matching works better than Integrate and Matrix determinant. We could conclude that string matching is more adapted for the applications with more methods and more complicated service graphs so that Ant Colony Optimization algorithm needs more time to evaluate a suitable path in it. In such complex applications, cache invalidation may also be more useful than for simple applications.

\section{Related Work}
\label{section:stateoftheart}

The idea of offloading is not a new concept~\cite{Balan:2003}, \cite{Balan:2002}, but it recently attracted much attention as a technique to overcome smartphone battery issues by partitioning and executing mobile applications on cloud-based surrogates. A significant amount of research has been performed to propose solutions to bring the cloud to the vicinity of a smartphone~\cite{Gi:2009}, \cite{Gordon2012}, \cite{Newton:2009}, \cite{Bo2012}.

In this section, we present a brief history of existing approaches with a focus on their structural aspects of offloading and decisions making mechanisms. In general, the offloading middleware makes either coarse-grain offloading as VM migration or fine-grained offloading concerning methods, jobs, classes, bundles, etc.
In both cases, they apply a single-objective decision-making process to decide what to offload, either statical at development or dynamical at runtime. Such a decision-making mechanism for single criteria leads to an optimal solution while our proposed bi-objective heuristic approach results in a trade-off between two criteria.

One of the most prominent works in this domain is MAUI~\cite{Cuervo:2010}. MAUI is an energy-aware offloading framework that uses developer code annotations to determine online which methods from a class must be offloaded if the bandwidth of the network and the data transfer conditions are ideal. However, MAUI does not address issues of adapting the mobile application for different devices and does not make advantage of the scalability feature of the cloud.
ThinkAir~\cite{Kosta2012} is similar to MAUI in that it provides method-level, semi-automatic offloading of code. However, ThinkAir focuses more on scalability issues and parallel execution of the offloaded tasks. It targets a commercial cloud scenario with multiple mobile users instead of computation offloading of a single user. Moreover, ThinkAir provides an efficient way to perform on-demand resource allocation and exploits parallelism by dynamically creating, resuming, and destroying VMs in the cloud when needed. However, since the development of mobile application uses annotations, the developer must follow a brute-force approach to adapt his/her application to a particular device.
Unlike MAUI and ThinkAir, in our work, an automated process defines remotable methods to eliminate developer burden in the development phase for method annotations. Also, fine-grained method-level application partitioning in ACOMMA makes lighter offloading compared with VM migration in these two middlewares.
CloneCloud~\cite{Chun:2011} is another middleware that profits from fine-grained application offloading, however its static analysis prevents from taking offloading decisions concerning the current situation. To ensure that mobile operations could be processed locally or remotely at bytecode level, CloneCloud also proposes the encapsulation of the mobile application as a stack into a virtual machine running in the cloud.
Odessa~\cite{Ra2011} takes dynamic offloading decisions based on an optimal solution during application partitioning. Unlike most of the middlewares that used linear programming to find such an optimal solutions, Odessa employs a greedy algorithm to this end.

\section{Conclusion}
\label{section:conclusion}

In this work, we focus on the mobile application augmentation by offloading to remote resources in the context of Mobile Cloud Computing. We introduce ACOMMA, an Ant-inspired Collaborative Offloading Middleware for Mobile Applications - a fine-grained method-level application offloading. Offloading decisions use a bi-Objective Ant Colony Optimization algorithm with execution time and CPU usage as criteria. To avoid running Ant Colony Optimization algorithm for any application offloading requests, we propose a learning-based decision-making model that searches an already taken decision in similar situations in previous execution trails and applies it to the current offloading process. ACOMMA works greatly to offload the applications that are more complex and resource-consuming. The more the application is consuming the more performance is augmented. This gain is high enough - 95 to 97\% of success with 75 to 99\% of gain - to consider the 5-7\% overhead of ACOMMA as acceptable.




\bibliographystyle{IEEEtran}
\bibliography{NOTERE-AntInspiredMobileOffloading}

\begin{thebibliography}{10}
\providecommand{\url}[1]{#1}
\csname url@samestyle\endcsname
\providecommand{\newblock}{\relax}
\providecommand{\bibinfo}[2]{#2}
\providecommand{\BIBentrySTDinterwordspacing}{\spaceskip=0pt\relax}
\providecommand{\BIBentryALTinterwordstretchfactor}{4}
\providecommand{\BIBentryALTinterwordspacing}{\spaceskip=\fontdimen2\font plus
\BIBentryALTinterwordstretchfactor\fontdimen3\font minus
  \fontdimen4\font\relax}
\providecommand{\BIBforeignlanguage}[2]{{%
\expandafter\ifx\csname l@#1\endcsname\relax
\typeout{** WARNING: IEEEtran.bst: No hyphenation pattern has been}%
\typeout{** loaded for the language `#1'. Using the pattern for}%
\typeout{** the default language instead.}%
\else
\language=\csname l@#1\endcsname
\fi
#2}}
\providecommand{\BIBdecl}{\relax}
\BIBdecl

\bibitem{Cis}
Cisco, ``{Cisco Visual Networking Index: Global Mobile Data Traffic Forecast
  Update},''
  \url{http://www.cisco.com/c/en/us/solutions/collateral/service-provider/
  visual-networking-index-vni/white\_paper\_c11-520862.html}, 2014.

\bibitem{Fernando2013}
N.~Fernando, S.~W. Loke, and W.~Rahayu, ``Mobile cloud computing: A survey,''
  \emph{Future Generation Computer Systems}, vol.~29, no.~1, pp. 84 -- 106,
  2013.

\bibitem{Cuervo:2010}
E.~Cuervo, A.~Balasubramanian, D.-k. Cho, A.~Wolman, S.~Saroiu, R.~Chandra, and
  P.~Bahl, ``{MAUI}: Making smartphones last longer with code offload,'' in
  \emph{Proc. of MobiSys '10}, 2010, pp. 49--62.

\bibitem{Kosta2012}
S.~Kosta, A.~Aucinas, P.~Hui, R.~Mortier, and X.~Zhang, ``Thinkair: Dynamic
  resource allocation and parallel execution in the cloud for mobile code
  offloading,'' in \emph{Proc. of IEEE INFOCOM '12}, March 2012, pp. 945--953.

\bibitem{Ghoseiri2010}
K.~Ghoseiri and B.~Nadjari, ``An ant colony optimization algorithm for the
  bi-objective shortest path problem,'' \emph{Appl. Soft Comput.}, vol.~10,
  no.~4, pp. 1237--1246, Sep. 2010.

\bibitem{face}
``Face recognition algorithm.'' [Online]. \url{Available:
  https://code.google.com/p/javafaces/}.

\bibitem{Balan:2003}
R.~K. Balan, M.~Satyanarayanan, S.~Y. Park, and T.~Okoshi, ``Tactics-based
  remote execution for mobile computing,'' in \emph{Proc. of MobiSys '03},
  2003, pp. 273--286.

\bibitem{Balan:2002}
R.~Balan, J.~Flinn, M.~Satyanarayanan, S.~Sinnamohideen, and H.-I. Yang, ``The
  case for cyber foraging,'' in \emph{Proc. of ACM SIGOPS EW '10}, 2002, pp.
  87--92.

\bibitem{Gi:2009}
I.~Giurgiu, O.~Riva, D.~Juric, I.~Krivulev, and G.~Alonso, ``Calling the cloud:
  Enabling mobile phones as interfaces to cloud applications,'' in \emph{Proc.
  of Middleware '09}, 2009, vol. 5896, pp. 83--102.

\bibitem{Gordon2012}
M.~S. Gordon, D.~Anoushe, J.~Scott, M.~Z. Morley, and M.~X. Chen, ``Comet: Code
  offload by migrating execution transparently,'' in \emph{Proc. of OSDI '12},
  2012, pp. 93--106.

\bibitem{Newton:2009}
R.~Newton, S.~Toledo, L.~Girod, H.~Balakrishnan, and S.~Madden, ``Wishbone:
  Profile-based partitioning for sensornet applications,'' in \emph{Proc. of
  NSDI '09}, 2009, pp. 395--408.

\bibitem{Bo2012}
B.~Gao, L.~He, L.~Liu, K.~Li, and S.~Jarvis, ``From mobiles to clouds:
  Developing energy-aware offloading strategies for workflows,'' in \emph{Proc.
  of GRID '12}, Sept 2012, pp. 139--146.

\bibitem{Chun:2011}
B.-G. Chun, S.~Ihm, P.~Maniatis, M.~Naik, and A.~Patti, ``{CloneCloud}: Elastic
  execution between mobile device and cloud,'' in \emph{Proc. of EuroSys '11},
  2011, pp. 301--314.

\bibitem{Ra2011}
M.-R. Ra, A.~Sheth, L.~Mummert, P.~Pillai, D.~Wetherall, and R.~Govindan,
  ``Odessa: Enabling interactive perception applications on mobile devices,''
  in \emph{Proc. of MobiSys '11}, 2011, pp. 43--56.

\end{thebibliography}

\end{document}